\documentclass[twocolumn,preprintnumbers,showpacs,prl,floatfix,amsmath,amssymb,notitlepage,superscriptaddress]{revtex4-1}
\usepackage{epsfig}
\usepackage{subfigure}
\usepackage{amsmath}
\usepackage{color}
\usepackage{amssymb}
\usepackage{setspace}
\usepackage{graphicx}
\usepackage{dcolumn}
\usepackage{bm}
\usepackage{times}
\usepackage{enumerate}
\usepackage{footnote}
\input epsf

\graphicspath{{figures/}}

\begin{document}

\author{Per Sebastian Skardal}
\email{skardals@gmail.com} 
\affiliation{Departament d'Enginyeria Informatica i Matem\'{a}tiques, Universitat Rovira i Virgili, 43007 Tarragona, Spain}
\affiliation{Department of Applied Mathematics, University of Colorado at Boulder, Colorado 80309, USA}

\author{Dane Taylor}
\email{dane.r.taylor@gmail.com}
\affiliation{Department of Applied Mathematics, University of Colorado at Boulder, Colorado 80309, USA}
\affiliation{Statistical and Applied Mathematical Sciences Institute, Research Triangle Park, NC 27709, USA}
\affiliation{Department of Mathematics, University of North Carolina, Chapel Hill, NC 27599, USA}

\author{Jie Sun}
\email{sunj@clarkson.edu}
\affiliation{Department of Mathematics, Clarkson University, Potsdam, NY 13699, USA}

\title{Optimal synchronization of complex networks}


\begin{abstract}
We study optimal synchronization in networks of heterogeneous phase oscillators. Our main result is the derivation of a {\it synchrony alignment function} that encodes the interplay between network structure and oscillators' frequencies and can be readily optimized. We highlight its utility in two general problems: constrained frequency allocation and network design. In general, we find that synchronization is promoted by strong alignments between frequencies and the dominant Laplacian eigenvectors, as well as a matching between the heterogeneity of frequencies and network structure.
\end{abstract}

\pacs{05.45.Xt, 89.75.Hc}

\maketitle

A central goal of complexity theory is to understand the emergence of collective behavior in large ensembles of interacting dynamical systems. Synchronization of network-coupled oscillators has served as a paradigm for understanding emergence \cite{Strogatz2003,Pikovsky2003,Dorogovtsev2008RMP,Arenas2008PR}, where examples arise in nature (e.g., flashing of fireflies~\cite{Buck2988QRB} and cardiac pacemaker cells~\cite{Glass1988}), engineering (e.g., power grid \cite{Motter2013NaturePhysics} and bridge oscillations~\cite{Strogatz2005Nature}), and at their intersection (e.g., synthetic cell engineering~\cite{Prindle2012Nature}). We consider the dynamics of $N$ network-coupled phase oscillators $\theta_i$ for $i=1,\dots,N$, whose evolution is governed by
\begin{align}
\dot{\theta}_i=\omega_i+K\sum_{j=1}^NA_{ij}H\left(\theta_j-\theta_i\right).\label{eq:Kuramoto}
\end{align}
Here $\omega_i$ is the natural frequency of oscillator $i$, $K>0$ is the coupling strength, $\left[A_{ij}\right]$ is a symmetric network adjacency matrix, and $H$ is a $2\pi$-periodic coupling function~\cite{Kuramoto1984}. We treat $H(\theta)$ with full generality so long as $H'(0)>0$. 
The choices $H(\theta)=\sin(\theta)$ and $H(\theta)=\sin(\theta-\alpha)$ with the phase-lag parameter $\alpha\in(-\pi/2,\pi/2)$ yield the classical Kuramoto \cite{Kuramoto1984} and Sakaguchi-Kuramoto models~\cite{Sakaguchi1986PTP}. 

Considerable research has shown that the underlying structure of a network plays a crucial role in determining synchronization~\cite{Moreno2004EPL,Ichinomiya2004PRE,Restrepo2005PRE,Arenas2006PRL,GomezGardenes2007PRL,Restrepo2007PRE,Gardenes2011PRL,Skardal2012PRE,Skardal2012Nolta,Skardal2013EPL}, yet the precise relationship between the dynamical and structural properties of a network and its synchronization remains not fully understood. One unanswered question is, given an objective measure of synchronization, how can synchronization be {\it optimized}? One application lies in synchronizing the power grid~\cite{Dorfler2013PNAS}, where sources and loads can be modeled as oscillators with different frequencies. To this end, we ask: what structural and/or dynamical properties should be present to optimize synchronization?

We measure the degree of synchronization of an ensemble of oscillators using the Kuramoto order parameter
\begin{align}
re^{i\psi} = \frac{1}{N}\sum_{j=1}^Ne^{i\theta_j}.\label{eq:OrderParameter}
\end{align}
Here $re^{i\psi}$ denotes the phases' centroid on the complex unit circle, with the magnitude $r$ ranging from $0$ (incoherence) to $1$ (perfect synchronization)~\cite{Kuramoto1984}. In general, the question of optimization (maximizing $r$) is challenging due to the fact that the macroscopic dynamics depend on both the natural frequencies and the network structure. To quantify the interplay between node dynamics and network structure, we derive directly from Eqs.~(\ref{eq:Kuramoto}) and (\ref{eq:OrderParameter}) a {\it synchrony alignment function} which is an objective measure of synchronization and can be used to systematically optimize a network's synchronization. We highlight this result by addressing two classes of optimization problem, which can be easily adapted to a wide range of applications. The first is {\it constrained frequency allocation}, where given a fixed network topology, optimal frequencies are chosen. The second is {\it network design}, where given a fixed set of frequencies, an optimal network structure is found. We next present the derivation of the synchrony alignment function. No assumptions are made about the frequencies or network aside from the network being connected and undirected.

We begin by considering the dynamics of Eq.~(\ref{eq:Kuramoto}) in the strong coupling regime where $r\approx1$, which may typically be obtained by either increasing the coupling strength or decreasing the heterogeneity of the frequencies. In this regime the oscillators are entrained in a tight cluster such that $\theta_i\approx\theta_j$ for all $(i,j)$ pairs. Expanding Eq.~(\ref{eq:Kuramoto}) yields
\begin{align}
\dot{\theta}_i\approx\tilde{\omega}_i-KH'(0)\sum_{j=1}^NL_{ij}\theta_j,\label{eq:Approx}
\end{align}
where $\tilde{\omega}_i=\omega_i+KH(0)d_i$, $d_i=\sum_{j=1}^NA_{ij}$ is the degree of node $i$, and $[L_{ij}]$ is the Laplacian matrix whose entries are defined as $L_{ij}=\delta_{ij}d_i-A_{ij}.$

The following spectral properties of the Laplacian are essential to our analysis. First, since the network is connected and undirected, all eigenvalues are real and can be ordered $0=\lambda_1 <\lambda_2\le\dots\le\lambda_{N-1}\le\lambda_N$. Second, the normalized eigenvectors $\{\bm{v}^i\}_{i=1}^N$ form an orthonormal basis for $\mathbb{R}^N$. Furthermore, the eigenvector associated with $\lambda_1=0$ is $\bm{v}^1=\bm{1}$, which corresponds to the synchronization manifold.

Inspecting Eq.~(\ref{eq:Approx}), we find that if a steady-state solution $\bm{\theta}^*$ exists after entering the rotating frame $\theta_i\mapsto\theta_i+\Omega t$, where $\Omega$ is the mean $\langle\tilde{\omega}\rangle$, it is given by
\begin{align}
\bm{\theta}^* = L^\dagger\bm{\tilde{\omega}}/KH'(0),\label{eq:stationarySolution}
\end{align}
where $L^\dagger=\sum_{j=2}^N\lambda_j^{-1}\bm{v}^j\bm{v}^{j T}$ is the pseudo-inverse of $L$~\cite{BenIsrael1974}. Under the approximation in Eq.~\eqref{eq:Approx}, the steady-state solution is expected to be linearly stable since the Jacobian matrix is approximately given by $-KH'(0)L$ and has nonpositive eigenvalues. We next consider the order parameter given $\bm{\theta}^*$. First, with a suitable shift in initial conditions the average phase can be set to zero, implying that the sum in Eq.~(\ref{eq:OrderParameter}) is real. Furthermore, in the strongly synchronized regime all phases are tightly packed about $\psi=0$, thus $|\theta_j^*|\ll1$ for all $j$. Expanding Eq.~(\ref{eq:OrderParameter}) 
yields
\begin{align}
r \approx1-\|\bm{\theta}^*\|^2/2N.\label{eq:rNorm}
\end{align}
Finally, by the spectral decomposition of the pseudo-inverse $L^\dagger$ and writing the norm in Eq.~(\ref{eq:rNorm}) by taking the inner product of $\bm{\theta}^*$ in Eq.~(\ref{eq:stationarySolution}), we obtain
\begin{align}
r=1-J(\bm{\tilde{\omega}},L)/2K^2H'^2(0),\label{eq:rObjective}
\end{align}
for which we define the synchrony alignment function
\begin{align}
J(\bm{\tilde{\omega}},L)=\frac{1}{N}\sum_{j=2}^N\lambda_j^{-2}\langle\bm{v}^j,\bm{\tilde{\omega}}\rangle^2.\label{eq:Objective}
\end{align}
The derivation of $J(\bm{\tilde{\omega}},L)$ is our main theoretical result as its minimization corresponds to the maximization of the order parameter $r$, which allows for the optimization of synchronization using elementary properties of the network (Laplacian eigenvalues and eigenvectors) and the frequencies. Before exploring the optimization of $J(\bm{\tilde\omega},L)$, we note the following interesting results. For $H(0)=0$, it follows that $\bm{\tilde{\omega}}=\bm{\omega}$, and thus optimization of $J(\bm{\omega},L)$ is independent of $K$. However, for $H(0)\ne0$ perfect synchrony, $r=1$, is generally not attainable unless $J(\bm{d},L)=0$ (which can occur if $d_1=d_2=\dots=d_N$)  since in the limit $K\to\infty$ Eq.~(\ref{eq:rObjective}) yields $r=1-J(\bm{d},L)H^2(0)/2H'^2(0)$. It follows that the existence of a strong coupling regime $r\lesssim1$ and consequently the approximations in our theory are valid only when $J(\bm{d},L)H^2(0)/2H'^2(0)\ll1$. Furthermore, since $\bm{\tilde{\omega}}$ depends on the coupling strength $K$, so will the optimization of $J(\bm{\tilde\omega},L)$ and therefore the optimal network. From now on, we will specialize to the widely used Kuramoto model, $H(\theta)=\sin(\theta)$, although we emphasize that similar results are found for more general coupling functions $H(\theta)$.

\begin{figure}[t]
\centering
\epsfig{file =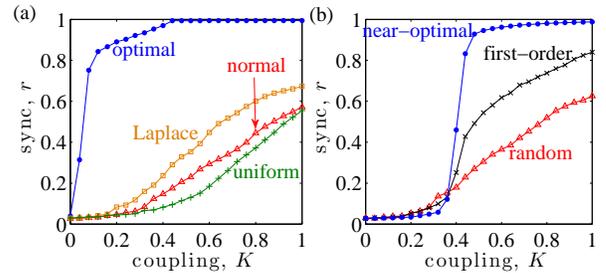, clip =,width=0.92\linewidth } 
\caption{(Color online) Constrained frequency allocation: (a) $r$ vs $K$ for optimal allocation (blue circles) compared with random allocations drawn from normal (red triangles), uniform (green pluses), and Laplace (orange squares) distributions. (b) $r$ vs $K$ for a pre-chosen set of normally distributed frequencies with random (red triangles), first-order (black crosses), and near-optimal (blue circles) allocations. The near-optimal allocation was obtained from $S=10^6$ proposed frequency exchanges. Networks are SF with $N=1000$, $\gamma=3$, and $d_0=2$.
} \label{fig:Optimal}
\end{figure}

We first address constrained frequency allocation for a fixed network. We note that by entering a rotating frame we can without loss of generality set the mean frequency to zero. The choice $\bm \omega=[0,\dots,0]^T$ trivially minimizes Eq.~\eqref{eq:Objective}, resulting in $r=1$, so we require as a first constraint that $\bf \omega$ has a fixed standard deviation, $\sigma=\sqrt{N^{-1}\sum_i\omega_i^2}$. By rescaling time and the coupling strength, $\sigma$ can be tuned freely, so without loss of generality we set $\sigma=1$. To minimize $J(\bm\omega,L)$, we first express $\bm{\omega}$ as a linear combination of the nontrivial eigenvectors of $L$, $\bm{\omega}=\sum_{i=2}^N\alpha_i\bm{v}^i$, where the coefficients must satisfy $\sum_{i=2}^N\alpha_i^2=N$. After inserting $\bm{\omega}$ into Eq.~(\ref{eq:Objective}) it follows that $J(\bm{\omega},L)$ is minimized by the choice $\alpha_2,\dots,\alpha_{N-1}=0$ and $\alpha_N=\sqrt{N}$, i.e., $\bm{\omega}\propto\bm{v}^N$, yielding $r=1-1/2\lambda_N^2K^2$.

In Fig.~\ref{fig:Optimal}~(a) we compare the results of optimal allocation, ${\bm \omega}=\sqrt{N}{\bm v}^N$, with several random frequency allocations by plotting the synchronization profiles $r$ vs $K$ for the optimal allocation, and those for frequencies randomly drawn from normal, uniform, and Laplace distribution (each with unit standard deviation). The underlying network with $N=1000$ nodes was constructed using the configuration model~\cite{Bekessy1972} with a scale-free (SF) degree distribution $P(d)\propto d^{-\gamma}$ for $\gamma=3$ and minimum degree $d_0=2$. The optimal allocation shows a large improvement over all random allocations and is marked by a sharp transition to a strongly synchronized state at a small coupling strength. In particular, the optimal allocation is surprisingly effective at very small coupling strengths despite the strong coupling assumption in the theory. We note that two mechanisms contribute to the excellent performance of the optimal allocation: the choice of frequencies and the nodes to which they are assigned.

To elucidate the importance of these two different mechanisms, we consider an additional constraint where frequencies $\{\omega_i\}_{i=1}^N$ are pre-chosen and must be allocated optimally on the network to minimize $J(\bm{\omega},L)$. Finding the global minimum requires an exhaustive search over all $N!$ possible permutations of $\bm{\omega}$ -- an unrealistic option even for moderately sized networks. We therefore provide two alternatives: a {\it first-order} approximation applicable for networks in which the largest Laplacian eigenvalue $\lambda_N$ is well separated from the others (often the case for SF networks \cite{Farkas2001}) and a {\it near-optimal} solution based on an accept/reject algorithm. In particular, when the dominant eigenvalue is well separated, $\lambda_i\ll\lambda_N$ for $i\not=N$, an inexpensive first-order minimization of the objective function leads to maximizing $|\langle\bm{v}^N,\bm{\omega}\rangle|$. This can be done simply by finding the index permutations $i_1,\dots,i_N$ and $j_1,\dots,j_N$ that place eigenvector entries in ascending order, $v_{i_1}^N\le\dots \le v_{i_N}^N$, and frequencies in ascending (or descending) order, $\omega_{j_1}\le\dots\le \omega_{j_N}$ (or $\omega_{j_1}\ge\dots\ge \omega_{j_N}$). In principle both pairings must be checked to select the best result. To find a near-optimal allocation we begin with an initial choice $\bm{\omega}$ and construct a new vector $\bm{\omega}'$ by exchanging two randomly chosen entries. If $J(\bm{\omega}',L) < J(\bm{\omega},L)$ we accept $\bm{\omega}'$, otherwise we reject it. This procedure is then repeated for $S$ proposed exchanges.

In Fig.~\ref{fig:Optimal}~(b) we compare synchronization profiles for near-optimal, first-order, and random allocations where frequencies are drawn from the unit normal distribution. As expected, the near-optimal allocation yields the best results, however, the first-order allocation also performs well, providing an inexpensive way to improve upon purely random allocation. These results also allow us to compare the allocation of pre-chosen frequencies to freely chosen frequencies [Fig.~\ref{fig:Optimal}~(a)]. In both cases, the transition from incoherence to strong synchronization is sharp, however it occurs at a larger coupling strength ($K\approx0.4$) when frequencies are pre-chosen, yielding two distinct regimes: for small $K$ strong synchronization is only attainable when frequencies are freely tunable, while for larger $K$ strong synchronization is attainable even when the frequency set is fixed. 

\begin{figure}[t]
\centering
\epsfig{file =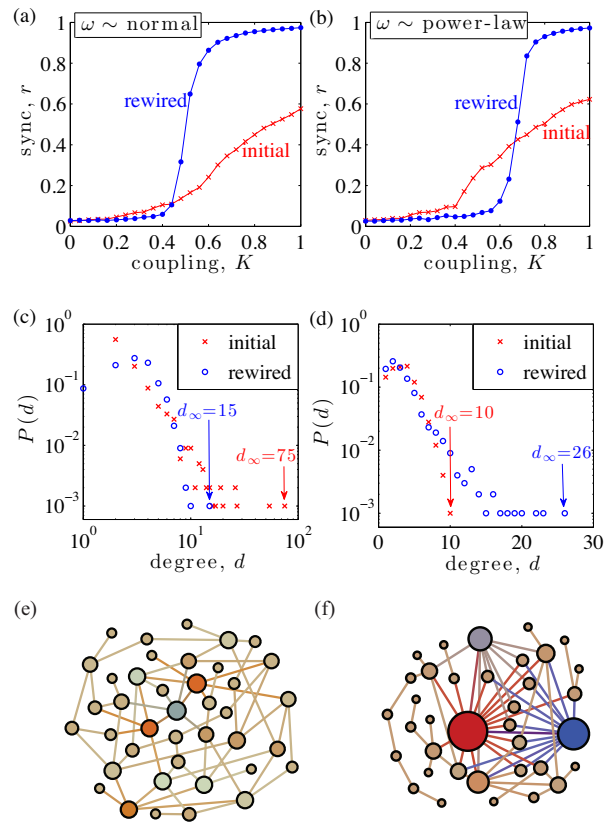, clip =,width=0.92\linewidth }
\caption{(Color online) Optimal network design: (a)-(b) $r$ vs $K$ for initial (red crosses) and rewired (blue circles) networks with normal and power-law distributed frequencies. (c)-(d) Degree distributions of the initial (red crosses) and rewired (blue circles) networks. (e)-(f) Illustrations for $N=40$ and $36$ of networks after rewiring.} \label{fig:Rewire}
\end{figure}

Next we address the complimentary problem of optimal network design for a fixed set of frequencies. Given $\bm{\omega}$ and a fixed number of links, we look for a network that minimizes $J(\bm\omega,L)$. As an algorithmic method for obtaining an approximate solution, we initialize an accept/reject algorithm with a network satisfying these constraints, and allow it to evolve as follows. A new network with Laplacian matrix $L'$ is constructed by randomly deleting a link and introducing another between two previously disconnected nodes. If $J(\bm{\omega},L')<J(\bm{\omega},L)$ we accept the new network, otherwise we reject it. This procedure is then repeated for $S$ proposed rewirings. In Fig.~\ref{fig:Rewire} we present the results of this rewiring algorithm for two experiments. We consider two networks: one with relatively homogeneous frequencies drawn from a unit normal distribution (left column) and a second with heterogeneous frequencies drawn from a symmetric power-law distribution, $g(|\omega|)\sim|\omega|^{-3}$ (right column). Both networks contain $N=1000$ oscillators. In Figs.~\ref{fig:Rewire}~(a) and (b) we plot the synchronization profiles for the initial networks and the networks obtained after $2\cdot10^4$ rewirings.  In both experiments, the rewired networks display better synchronization properties with sharp transitions from incoherence to strong synchronization. Each experiment is initialized with a different network topology: a SF network constructed by the configuration model with $\gamma = 3$ and $d_0=2$ and an Erd\H{o}s-R\'{e}nyi (ER)~\cite{Erdos1960} network with average degree $\langle d\rangle=4$ are paired with the normal and power-law distributed frequencies, respectively. In Figs.~\ref{fig:Rewire}~(c) and (d) we plot the initial and rewired degree distributions. In both experiments the degree distribution evolves to better match that of the frequencies, either becoming less [Fig.~\ref{fig:Rewire}(c)] or more [Fig.~\ref{fig:Rewire}(d)] heterogeneous. This is further emphasized by the shifts in the maximal degrees $d_\infty$, which decreases from 75 to 15 and increases from 10 to 26, respectively. This suggests that a more heterogeneous network better synchronizes a more heterogeneous set of frequencies. To illustrate this phenomenon, we show in Figs.~\ref{fig:Rewire}~(e) and (f) networks resulting from the same experiment with fewer nodes ($N=40$ and $36$, respectively). The radius of each node is proportional to its degree and the coloring of the node indicates its frequency from most positive (red) to most negative (blue). Here the phenomenon is easily observable with the emergence of network hubs in (f) but not (e). 

\begin{figure}[t]
\centering
\epsfig{file =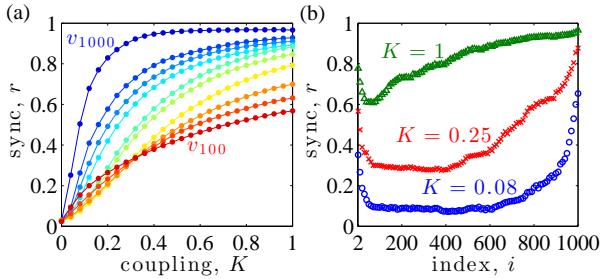, clip =,width=0.92\linewidth }
\caption{(Color online) Eigenvector alignments: (a) $r$ vs $K$ for frequency alignments $\bm{\omega}\propto\bm{v}^{100}, \bm{v}^{200},\dots,\bm{v}^{1000}$ (red to blue). (b) $r$ vs $i$ for $\bm{\omega}\propto\bm{v}^i$ with fixed $K=0.08$ (blue circles), $0.25$ (red crosses), and $1$ (green triangles). Each point is averaged over $50$ SF network realization of size $N=1000$ with $\gamma=3$ and $d_0=2$.} \label{fig:Vectors}
\end{figure}

We now study in more detail the synchrony alignment function given in Eq.~\eqref{eq:Objective}. Just as aligning $\bm{\omega}$ with $\bm{v}^N$ maximizes $r$, it follows that in the strong coupling regime, aligning $\bm{\omega}$ with other eigenvectors $\bm{v}^i$ of decreasing index yields weaker synchronization. We consider the alignments $\bm{\omega}\propto\bm{v}^i$ and plot the synchronization profiles in Fig.~\ref{fig:Vectors}~(a) for $i=100,200,\dots,1000$ (red to blue) averaged over $50$ realizations of SF networks with parameters $N=1000$, $\gamma=3$, and $d_0=2$. As expected, we observe weaker synchronization with decreasing index. We also plot in panel (b) $r$ vs $i$ for a few isolated coupling strengths: $K=0.08$, $0.25$, and $1$, again averaged over 50 realizations. For all three cases $r$ tends to increase with $i$, provided $i$ is not too small. For $K=0.08$ the majority of alignments yield incoherence with $r$ undergoing a sharp increase only near the most dominant eigenvectors, while the increase in $r$ is more gradual for $K=0.25$ and $1$. We also point out that for very small $i$ we observe a short increase in $r$, which we attribute to local synchronization that yield large fluctuations in $r$.

\begin{figure}[t]
\centering
\epsfig{file =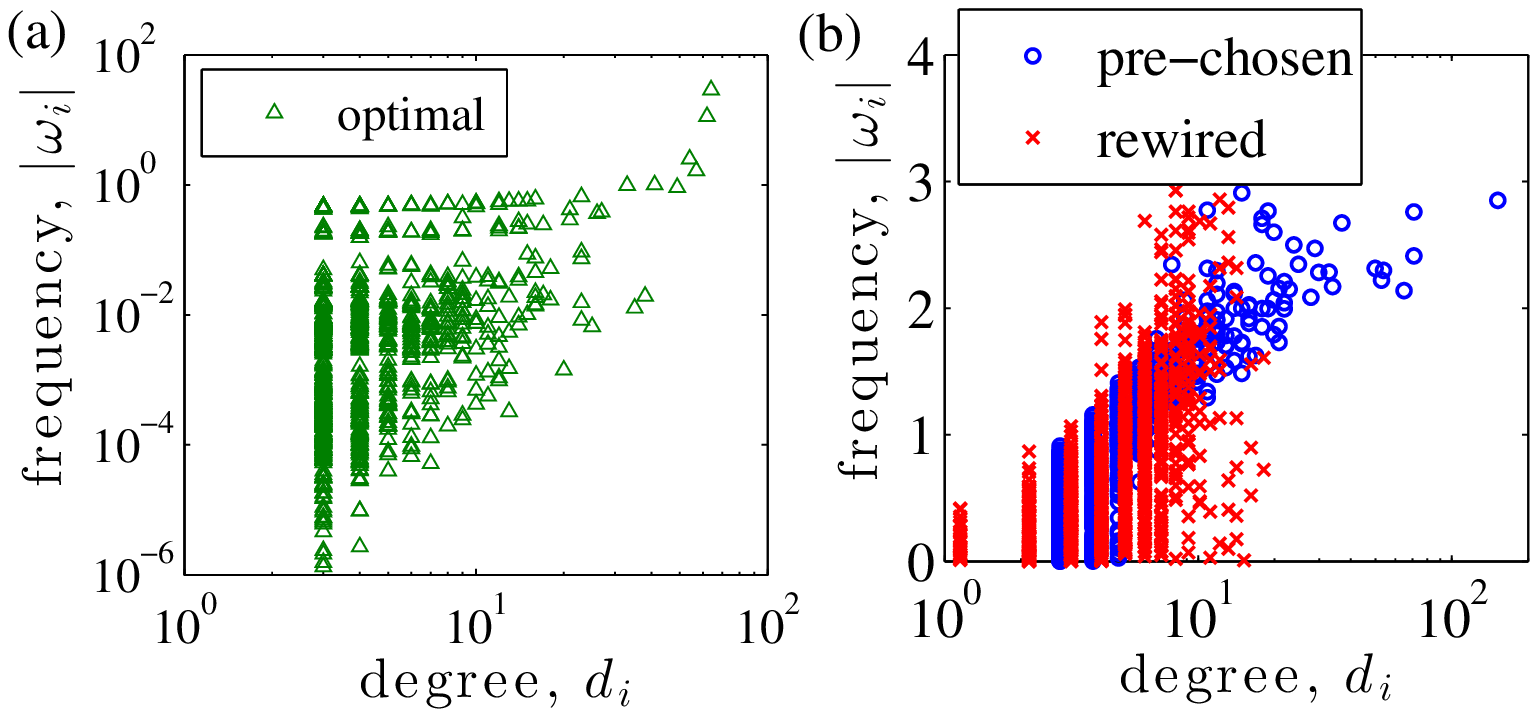, clip =,width=0.92\linewidth }\\
\epsfig{file =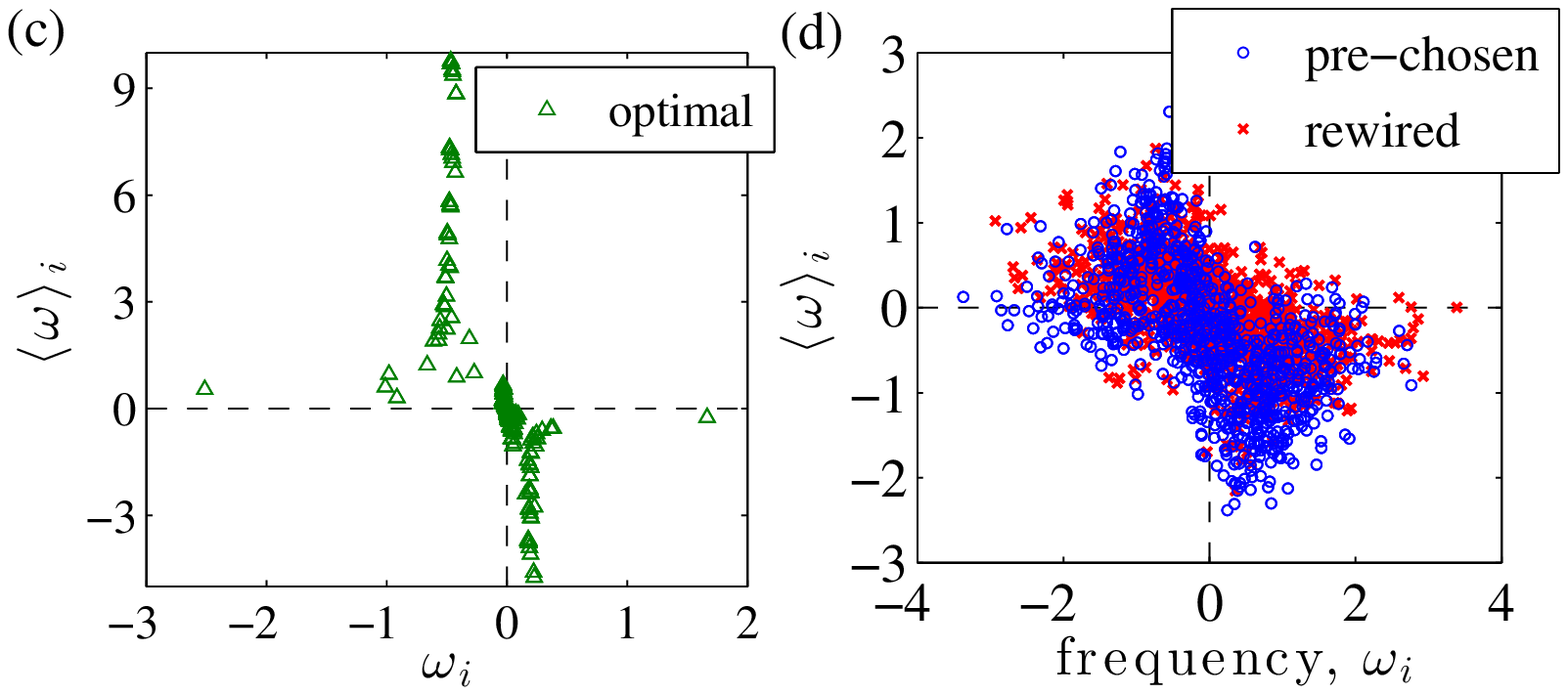, clip =,width=0.92\linewidth }
\caption{(Color online) Correlations in optimized networks: (a), (b) Frequency magnitude $|\omega_i|$ vs degree $d_i$ and (c), (d) average neighbor frequency $\langle\omega_i\rangle$ vs frequency $\omega_i$ for networks with optimally chosen frequencies (green triangles), pre-chosen frequencies (blue circles), and a rewired network (red crosses). Networks are SF with $N=1000$, $\gamma=3$, and $d_0=3$. Frequencies for the arranged and rewired cases are normally distributed.} \label{fig:Correlations}
\end{figure}

Before concluding, we investigate the dynamical and structural properties present in optimized networks. In particular, we consider local degree-frequency and neighboring frequency-frequency correlations. In Figs.~\ref{fig:Correlations}~(a) and (b) we plot the frequency magnitude $|\omega_i|$ vs degree $d_i$ for (a) a network with optimally allocated frequencies and (b) a network with pre-chosen frequencies (blue circles) and a rewired network (red crosses). Networks are SF with $N=1000$, $\gamma=3$, and $d_0=3$ and frequencies in (b) are normally-distributed. In each case we observe a strong positive degree-frequency correlation, indicating that the largest frequencies correspond to the network hubs. Moreover, in Figs.~\ref{fig:Correlations}~(c) and (d) we plot for each node $i$ the average frequency of its neighboring oscillators $\langle\omega\rangle_i=\sum_{j=1}^NA_{ij}\omega_j/d_i$ vs $\omega_i$. The results are qualitatively similar for each case, showing a strong negative correlation between neighboring frequencies. These observations agree with those of Refs.~\cite{Brede2008PLA,Buzna2009PRE}, where similar positive and negative correlations were found to promote global synchronization. We finally note that such degree-frequency correlations may help explain the increased sharpness of transitions shown in Figs.~\ref{fig:Optimal} and \ref{fig:Rewire} (a) and (b), since similar correlations can lead to discontinuous transitions~\cite{Gardenes2011PRL}.

In this Letter we presented a {\it synchrony alignment function} that measures the interplay between network structure and oscillator heterogeneity and allows for a systematic optimization of synchronization. Focusing on Kuramoto coupling, we highlighted its utility through numerical experiments for random networks with two general classes of optimization problems: frequency allocation and network design. We found that synchronization is promoted by a strong alignment of the frequency vector with the most dominant Laplacian eigenvectors and that, relatively speaking, more (less) heterogeneous networks better synchronize more (less) heterogenous frequencies. In all cases we found that in optimized networks the large frequencies are localized to hubs and frequencies of neighboring oscillators are negatively correlated.

Although the theoretic approach developed herein is valid for systems given by Eq.~\eqref{eq:Kuramoto}, extension to more general oscillator models, e.g., Landau-Stuart oscillators~\cite{Rosenblum2007PRL}, Winfree oscillators~\cite{Winfree1967JTB} and chaotic oscillators remains an outstanding problem. One promise stems from Kuramoto's phase reduction methods which give Eq.~\eqref{eq:Kuramoto} as an approximating of the dynamics of weakly-interacting limit-cycle oscillators~\cite{Kuramoto1984}. Another exciting venue of research would be on the optimization of other dynamical patterns such as multistability, hysteresis, and/or explosive synchronization, none of which was observed in our numerical examples (despite the sharp transitions seen in Figs.~\ref{fig:Optimal} and~\ref{fig:Rewire}) but can potentially arise under more general coupling and dynamics.

Finally, we compare our results on heterogeneous oscillators to the well-developed theory regarding identical oscillators~\cite{Pecora1998PRL}, for which the synchronizability of a network is given by the ratio $\lambda_N/\lambda_2$ of  Laplacian eigenvalues~\cite{Barahona2002PRL} -- a result allowing for optimization to be independent of the node dynamics~\cite{Nishikawa2010PNAS}. In contrast, we find here that the synchronization of a network of heterogeneous oscillators generally depends on not only the full set of eigenvalues and eigenvectors of the network Laplacian, much like the case of nearly-identical oscillators~\cite{Sun2009EPL} and real-world experiments~\cite{Ravoori2011PRL}, and how the network structure pairs with the heterogeneity of node dynamics (here oscillator frequencies). A network that is easily synchronizable with identical oscillators may have poor synchronization properties with heterogeneous oscillators, and vice-versa.

\acknowledgements
This work was funded in part by the James S. McDonnell Foundation (PSS), NSF Grant No.~DMS-1127914 through the Statistical and Applied Mathematical Sciences Institute (DT), ARO Grant No.~61386-EG (JS), and Simons Foundation Grant No.~318812 (JS).

\bibliographystyle{plain}

\end{document}